\documentclass[%
superscriptaddress,
notitlepage,
 amsmath,
 amssymb,
 twocolumn,
 aps,
prb,
]{revtex4-1}

\usepackage{graphicx}
\usepackage{dcolumn}
\usepackage{bm}

\begin{document}

\title{Magnetophonon spectroscopy of Dirac Fermion scattering by transverse and longitudinal acoustic phonons in graphene}

\affiliation{Department of Physics, Loughborough University, LE11 3TU, UK}

\affiliation{School of Physics and Astronomy, University of Nottingham, NG7 2RD, UK}

\affiliation{School of Physics and Astronomy, University of Manchester, Manchester, M13 9PL, UK}

\affiliation{National Graphene Institute, University of Manchester, Manchester, M13 9PL, UK}

\author{M.T. Greenaway}
\email{m.t.greenaway@lboro.ac.uk}
\affiliation{Department of Physics, Loughborough University, LE11 3TU, UK}
\affiliation{School of Physics and Astronomy, University of Nottingham, NG7 2RD, UK}

\author{R. Krishna Kumar}
\affiliation{School of Physics and Astronomy, University of Manchester, Manchester, M13 9PL, UK}

\author{P. Kumaravadivel}
\affiliation{School of Physics and Astronomy, University of Manchester, Manchester, M13 9PL, UK}
\affiliation{National Graphene Institute, University of Manchester, Manchester, M13 9PL, UK}

\author{A.K. Geim}
\affiliation{School of Physics and Astronomy, University of Manchester, Manchester, M13 9PL, UK}
\affiliation{National Graphene Institute, University of Manchester, Manchester, M13 9PL, UK}

\author{L. Eaves}
\email{laurence.eaves@nottingham.ac.uk}
\affiliation{School of Physics and Astronomy, University of Nottingham, NG7 2RD, UK}
\affiliation{School of Physics and Astronomy, University of Manchester, Manchester, M13 9PL, UK}

\date{\today}

\begin{abstract}

Recently observed magnetophonon resonances in the magnetoresistance of graphene are investigated using the Kubo formalism. This analysis provides a quantitative fit to the experimental data over a wide range of carrier densities. It demonstrates the predominance of carrier scattering by low energy transverse acoustic (TA) mode phonons: the magnetophonon resonance amplitude is significantly stronger for the TA modes than for the longitudinal acoustic (LA) modes. We demonstrate that the LA and TA phonon speeds and the electron-phonon coupling strengths determined from the magnetophonon resonance measurements also provide an excellent fit to the measured dependence of the resistivity at zero magnetic field over a temperature range of 4-150 K.  A semiclassical description of magnetophonon resonance in graphene is shown to provide a simple physical explanation for the dependence of the magneto-oscillation period on carrier density.  The correspondence between the quantum calculation and the semiclassical model is discussed.

\end{abstract}

\maketitle

\section{Introduction}

In 1961, theoretical work by Gurevich and Firsov predicted that inelastic scattering of electrons by phonons can induce oscillations in the magnetoresistance of semiconductors \cite{Gurevich1961}.  Magnetophonon resonance (MPR) has since been used to probe spectroscopically electron-phonon interactions in a wide range of bulk semiconductors \cite{Peterson1975,Firsov1964,Stradling1968,Eaves1975} and semiconductor heterostructures in which carriers are confined in two-dimensions (2D) by a quantum well potential \cite{Tsui1980,Nicholas1985,Zudov2001,Hatke2009,Dmitriev2012}.  

Early studies of MPR focused mostly on carrier scattering between Landau levels (LLs), induced by weakly dispersed longitudinal optical (LO) phonons with a well-defined energy, $\hbar \omega_{LO}$, and a high density of states.  The resonant condition is given by $\hbar \omega_{LO}=p\hbar \omega_c$, where $\omega_c=eB/m^{*}$ is the cyclotron frequency, $B$ is the applied magnetic field amplitude, $m^{*}$ is the carrier effective mass and $p$ is an integer.  Absorption or emission of a phonon can induce a shift of the electron's cyclotron orbit centre.  This causes it to drift in the presence of an applied voltage and give rise to an enhancement of the magnetoconductance when the resonant condition is satisfied.  The result is a series of oscillations in the magnetoconductance that are observable over a wide range of temperatures, are periodic in inverse magnetic field and are independent of carrier density.

A different type of MPR was observed at low temperatures in the magnetoresistance of a modulation doped (AlGa)As-GaAs heterostructure \cite{Zudov2001,Hatke2009,Dmitriev2012}.  Under these conditions MPR was shown to arise from scattering of the two-dimensionally confined electrons by linearly dispersed acoustic phonons. The oscillatory period, $\Delta(B^{-1})$, had a square root dependence on carrier sheet density. 

\begin{figure}[!t]
  \centering
\includegraphics[width=1\linewidth]{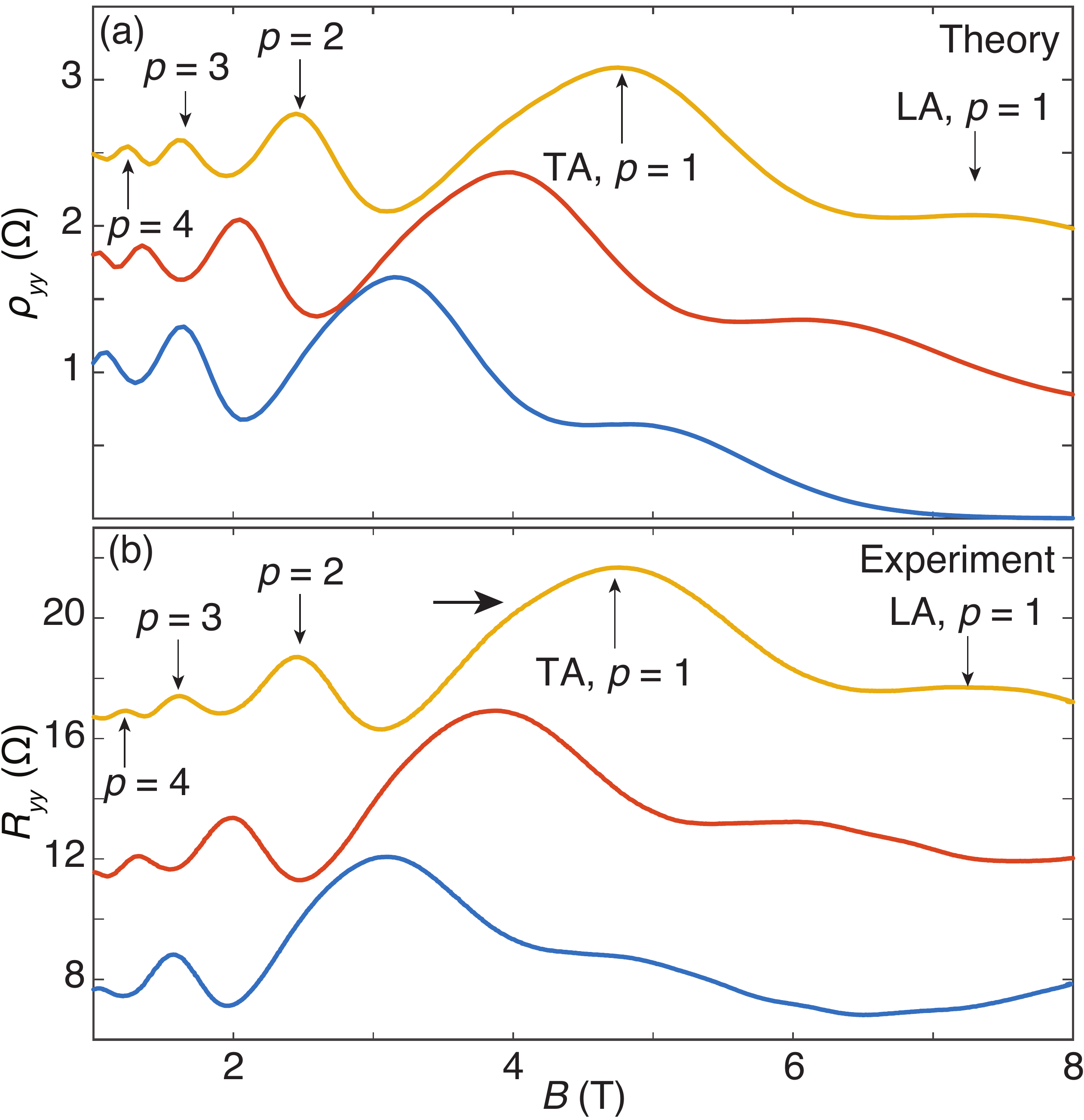}
  \caption{(a) Calculated, and (b), measured longitudinal magnetoresistivity, $\rho_{yy}(B)$ and  magnetoresistance $R_{yy}(B)$ in a 15 $\mu$m wide gated Hall bar, at $T$=70 K.  The blue, red and green curves correspond to carrier densities $n_s=6.0,$ 7.5 and 9.0 $\times 10^{12}$ cm$^{-2}$ respectively in both sets of plots.  Curves in (a) are offset by 0.75 $\Omega$ for clarity.}
\label{fig:ndep}
\end{figure}

Here, we present a theoretical model to investigate large amplitude acoustic phonon-induced magnetoresistance oscillations that were observed recently in wide, gated Hall bars of monolayer graphene encapsulated in hexagonal boron nitride (hBN) \cite{Kumaravadivel2019}. The spectroscopic nature of MPR complements the extensive literature on the effects of electron-phonon interactions on the carrier mobility of graphene \cite{Suzuura2002,Mariani2010,vonoppen2009,Sohier2014,Hwang2008,Borysenko2010,Castro2010,Park2014,Kaasbjerg2012,Efetov2010,Morozov2008,Alexeev2015,Manes2007,Wu2019,Yudhistira2019,Polshyn2019} and recent measurements of phonon-assisted tunnelling in stacked graphene-hBN-graphene devices \cite{Vdovin2016,Jung2015}. Fig. \ref{fig:ndep} compares the results of our calculation with the experimental data \cite{Kumaravadivel2019}.  The peaks in magnetoresistance are periodic in $1/B$ with a frequency, $B_F$, that is linearly dependent on the carrier density, $n_s$, of the Dirac Fermions.  The dependence of $B_F$ on $n_s$ allows us to determine the speeds of the linearly dispersed TA and LA phonons which give rise to the magneto-oscillations.  The analysis demonstrates that it is necessary to include scattering by both TA and LA phonons to obtain a quantitative understanding of graphene's phonon-limited resistivity.    We demonstrate how electrical screening of the deformation potential accounts, in part, for the smaller amplitude of the LA phonon resonances. 

\begin{figure}[!t]
  \centering
\includegraphics[width=1\linewidth]{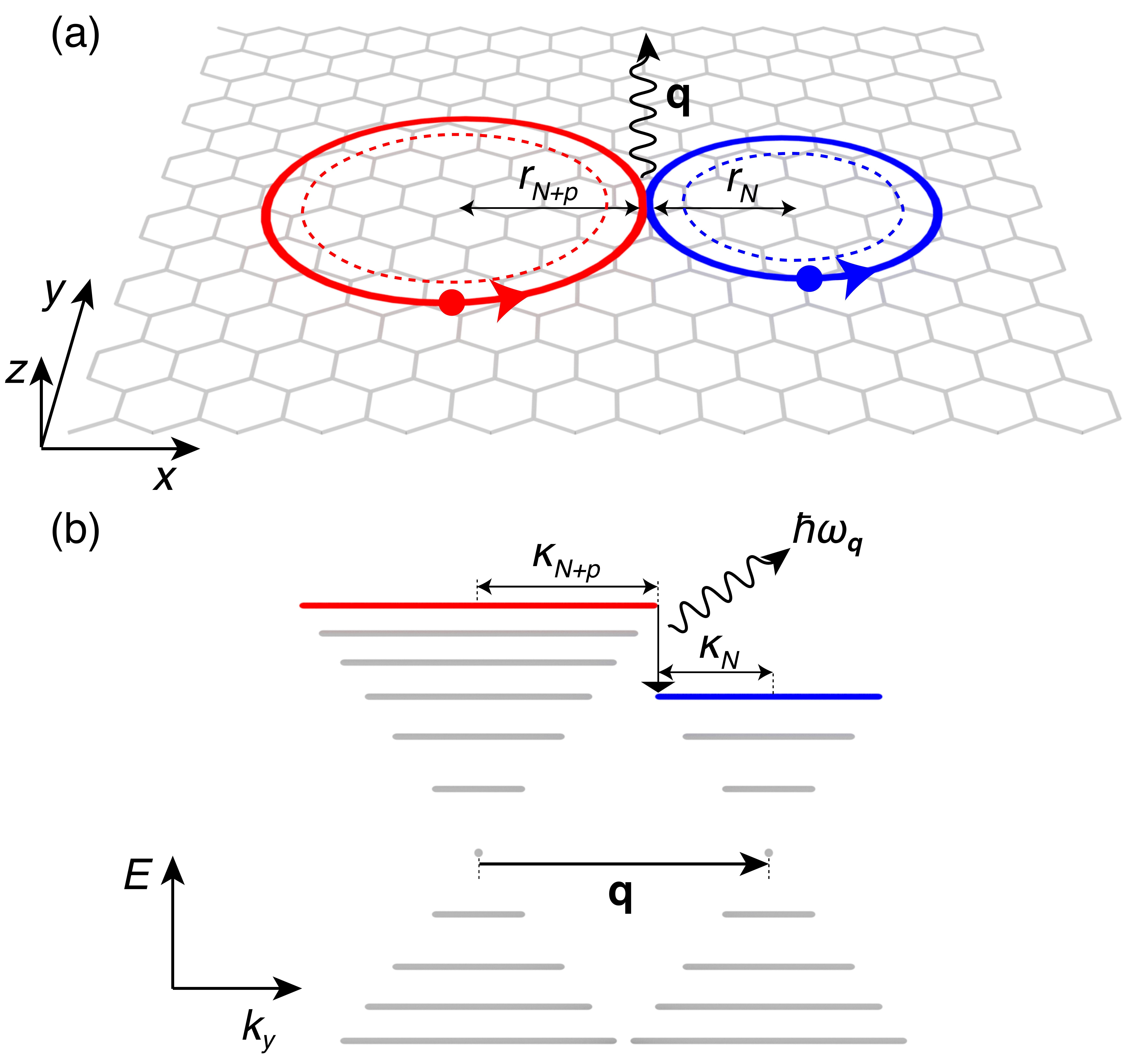}
  \caption{(a) Schematic diagram of the coordinate axes and scattering of a Dirac Fermion in the graphene lattice. Red and blue circles represent the real space cyclotron orbits, radii $r_{N+p}$ and $r_N$, of an electron (filled circle) before and after scattering by a phonon with wavevector ${\bf q}$ and the shift of its orbit centre.  The area between the dashed and full circles show schematically the width of the largest peak of the LL wavefunction adjacent to its classical turning point. (b) The horizontal lines show the energies and diameters of cyclotron orbits in $k$-space  before and after scattering by a phonon with wavevector {\bf q}.  Red and blue lines show magneto-acoustic-phonon resonance between an initial state with radius $\kappa_{N+p}$, and final state with radius $\kappa_N$.}
\label{fig:schematic}
\end{figure}

\section{Semiclassical analysis}
\label{sec:2}
Fig. \ref{fig:schematic}(a) shows a schematic diagram of the coordinate and lattice orientation used in our calculation.  The measurements in Fig. \ref{fig:ndep}(b),  were made with a negative gate voltage applied to the device so that the charge carriers are holes with Fermi energy, $E_F$, positioned below the Dirac point of graphene's band structure \cite{Kumaravadivel2019}.  Our analysis considers the case when $E_F$ is above the Dirac point;  electron-hole symmetry close to the Dirac point in graphene ensures that it is applicable to both types of charge carrier. 

Application of a magnetic field, ${\bf B}=(0,0,-B)$ where $B=|{\bf B}|$, perpendicular to the graphene sheet quantises the electron energy, $E_N$, into a series of unevenly spaced LLs with index $N$, given by the relation
\begin{equation}
E_N=\sqrt{2N}\frac{\hbar v_F}{l_B},
\label{eq:EN}
\end{equation}
where $v_F$ is the Fermi velocity in graphene and $l_B=\sqrt{\hbar/eB}$ is the magnetic length. 

In the absence of scattering, carriers would propagate freely in the direction perpendicular to an applied electric field so that the magnetoconductivity $\sigma_{xx}=0$.  When a carrier scatters inelastically by the emission or absorption of a phonon with wavevector ${\bf q}$, momentum conservation requires that its orbit centre shifts, giving rise to a dissipative current and finite $\sigma_{xx}$, and magnetoresistivity $\rho_{yy}$.   

The amplitude of the MPR oscillations increases with increasing temperature up to $\sim 100$ K.  Thermal excitation of phonons and broadening of the Fermi distribution enable a carrier at the Fermi energy $E_F$, to absorb or emit a phonon.  At higher temperatures, phonon scattering is sufficiently strong to prevent a carrier from completing a cyclotron orbit ($\mu B< 1$, where $\mu$ is the carrier mobility). The width of the LL then becomes comparable with the LL energy separation and MPR oscillations are damped out.  

Inelastic scattering between two LLs with indices differing by the integer $p=1,2,3...$ occurs when the acoustic phonon energy, $\hbar\omega_a$, equals the difference in their energies:
\begin{equation}
E_{N+p}-E_N=\pm\hbar\omega_a.
\label{eq:Econd}
\end{equation}
For small $q$, acoustic phonons have a linear dispersion relation given by 
\begin{equation}
\omega_q^a=v_a q,
\end{equation}
where $v_a$ is the phonon velocity.  In semiclassical Newtonian dynamics, a carrier with energy $E_N$ performs closed cyclotron orbits in real space with radius $r_N=l_B\sqrt{2N}$ and, since $\hbar {\bf \dot k}=- e{\bf v} \times {\bf B}$, in $k-$space with radius
\begin{equation}
\kappa_N=\frac{\sqrt{2N}}{l_B}.
\label{eq:k_n}
\end{equation}
Inelastic scattering by a phonon shifts the centre of the cyclotron orbit in $k$-space by ${\bf q}$ and its position in real space by $(\Delta X,\Delta Y)=l_B^2(-q_y,q_x)$. Semiclassically, this can occur only when the orbits of the initial and final states intersect.  The onset of inelastic scattering occurs when the cyclotron orbits just touch, giving rise to a trajectory with a ``figure-of-8'' orbit \cite{Greenaway2015}, see Fig. \ref{fig:schematic}. Thus
\begin{equation}
\kappa_{N+p}+\kappa_N=q.
\label{eq:kcond}
\end{equation}
To a good approximation, this semiclassical description corresponds to the condition for the maximum in the overlap of the wavefunctions of the initial and final states.  By combining Eqs. \ref{eq:Econd}-\ref{eq:kcond} we obtain the magnetophonon resonance condition:
\begin{equation}
v_F\left(\sqrt{N+p}-\sqrt{N}\right)= v_a \left(\sqrt{N+p}+\sqrt{N}\right),
\label{eq:MAPcondition}
\end{equation}
from which we obtain the following semiclassical equation for $N=N_p$, when Eq. \ref{eq:MAPcondition} is satisfied:
\begin{equation}
N_p=\frac{pv_a}{4v_F}\left(\frac{v_F}{v_a}-1\right)^2 \approx \frac{pv_F}{4v_a}.
\label{eq:m}
\end{equation}
The approximation is valid since $v_F\gg v_a$. Our calculation of the resonant scattering processes involves transitions between LLs with indices of up to $N\sim100$. The large energy separation between LLs with $N<N_p$ requires a high energy phonon with a $q$ that is too large to allow the semiclassical orbits of the initial and final to intersect; scattering cannot then occur.  As will be discussed in section \ref{sec:kubo} this condition is relaxed in quantum mechanics. At the classical turning point of the LL wavefunctions, each well-defined cyclotron orbit is effectively broadened into a ring of width $\sim l_B$, see dashed and full circles in Fig. \ref{fig:schematic}.

We develop the semiclassical model by considering electrons within a range of $\approx E_F\pm 2 k_B T$. Thus we set $E_{N_p}=\hbar v_F k_F$, where $k_F=\sqrt{\pi n_s}$ is the Fermi wavevector and $n_s$ is the carrier density.  We then obtain
\begin{equation}
N_p=\frac{l_B^2k_F^2}{2}=\frac{\hbar \pi n_s}{2eB_p},
\end{equation}
where $B_p$ is the magnetic field corresponding to a maximum in  $\sigma_{xx}$ and $\rho_{yy}$.  Using this expression in Eq. \ref{eq:m}, we obtain
\begin{equation}
B_p=\frac{n_s h v_F}{p e v_a} \left(\frac{v_F}{v_a}-1\right)^{-2}\approx \frac{n_s h v_a}{p e v_F} .
\label{eq:Blexact}
\end{equation}

This relation describes accurately the data shown in Fig. \ref{fig:ndep}(b) and described in Kumaravadivel {\it et al.} \cite{Kumaravadivel2019}.  The measured magneto-oscillations in Fig. \ref{fig:ndep}(b) reveal a strong set of peaks labelled ``TA, p=1,2...''  that are periodic in $B^{-1}$ with a well-defined frequency, $B_F=pB_p$, that is linearly dependent on $n_s$.  We associate these peaks with MPR due to TA phonons.  The dependence of $B_F$ on $n_s$ indicates a constant ratio between the speed of the TA acoustic phonon, $v_{TA}$, and the Fermi velocity so that $v_{TA}/v_F=0.0128$.   We also observe a weaker peak at higher $B$, labelled ``LA, p=1'', with a $B_F$ value that is linearly dependent on $n_s$.  We associate this peak with MPR due to LA phonons with speed $v_{LA}$ and we obtain $v_{LA}/v_F=0.0198$.  With a Fermi velocity of $v_F = 1.06 \pm  0.05 \times 10^6$ ms$^{-1}$ extracted from the temperature-dependent Shubnikov de Haas measurements on the devices reported by Kumaravadivel {\it et al.}\cite{Kumaravadivel2019}, we obtain $v_{TA} = 13.6\pm0.7$ kms$^{-1}$ and $v_{LA}= 21\pm1$ kms$^{-1}$.  These values are in good agreement with calculations of the speeds of linearly dispersed acoustic phonons in graphene \cite{Park2014,Sohier2014,Kaasbjerg2012}.  We note that the measured constant ratio between $v_a/v_F$ is fully consistent with the constancy of both $v_F$ and $v_a$ over the range of $n_s$ from $1.5$ to \mbox{ $9\times 10^{16}$ m$^{-2}$ } and $q$ from $\sim0.5$ to $1.0\times10^9$ m$^{-1}$.  A constant $v_F$ is expected in graphene devices on dielectric substrates over this range of $n_s$ \cite{Yu2013} due to screening of electron-electron interactions that cause velocity renormalization \cite{Elias2011}.

To conclude this section we compare relation \ref{eq:Blexact} with earlier work on 2DEGs in III-V heterostructures \cite{Zudov2001,Hatke2009} where electrons have a parabolic dispersion and well-defined effective mass, $m^*$. The energy separation between LLs, $\hbar\omega_c$, is then independent of $N$. In this case the MPR resonant condition is given by
\begin{equation}
\hbar \omega_c=\hbar v_a \left(\kappa_{N+p}+\kappa_{N}\right),
\end{equation}
so that
\begin{equation}
B^{\rm 2DEG}_p\approx \frac{2 m^*v_a k_F}{p e}= \frac{2m^* v_a \sqrt{2 \pi n_s}}{p e}.
\label{eq:Bp2DEG}
\end{equation}

This expression is similar to relation \ref{eq:Blexact} for graphene, in particular the oscillations are periodic in $1/B$ .  However, in contrast to graphene, the position of the resonant peak depends on the square root of the carrier density.

\section{A quantum calculation of $\rho_{yy}$}
\label{sec:kubo}
The semiclassical model in section \ref{sec:2} can be used to obtain the resonance condition but not the amplitude and shape of the oscillations.  We therefore present in this section a full quantum mechanical calculation of $\rho_{yy}$ based on the Kubo formalism \cite{Kubo1965,Hamaguchi1990,Mori2011}.  It is convenient to choose the Landau gauge where ${\bf A}=(0, -Bx, 0)$. The Dirac Fermion wavefunction in the $K^+$ valley for $|N|>0$ is given by the pseudospinor
\begin{equation}
\psi_{N,X}=\frac{1}{\sqrt{2}}\ \begin{pmatrix}\phi_{|N|}(x-X) \\ -\textrm{sgn}(N)i\phi_{|n|-1}(x-X)\end{pmatrix},
\end{equation}
where $\phi$ are simple harmonic oscillator states given by
\begin{equation}
\phi_N(x)=A_N H_N \left(\frac{x}{l_B}\right) \exp \left(-\frac{x^2}{2 l_B^2}\right) \exp\left(i k_y y\right).
\end{equation}
A similar expression applies to carriers in the $K^-$ valley.  Here, $A_N=1/\sqrt{L l_B 2^N N! \pi^{1/2}}$ is a normalisation constant, $H_N$ are the Hermite polynomials \cite{McCannbook,Shon1998,CastroNeto2009}, and $L$ is the size dimension of the Hall bar. With this choice of gauge, the wavefunctions can be thought of as a series of strips along the $y$-axis of the Hall bar centred on $X=l_B^2 k_y$ and comprising plane waves with wavevector $k_y$ along the $y$ and Hermite polynomials along $x$ \cite{Beenakker1991}. The magnetoconductance depends on the rate of drift of the orbit centre due to phonon scattering and is given by 
\begin{align}
\sigma_{xx}^a=\frac{g_v g_s\pi e^2}{L^4 k_B T \hbar} \sum_{\bf q} (l_B^2 q_y)^2 |C_a(q)|^2 N_q(N_q+1) \nonumber \\ \times \sum_{N,N'} \sum_{{k_y},{k_y'}} \left[ f(E_{N}-\hbar\omega_{q}^a) - f (E_{N}) \right]  \nonumber \\
\times \delta\left(E_{N} - \hbar\omega_q^a -E_{N'} \right) | I^a_{N,N'}({k_y},{k_y'},{\bf q})|^2
\label{eq:Kubo}
\end{align}
for the TA ($a=TA$) and LA ($a=LA$) phonons, where $g_v=2$ and $g_s=2$ are the valley and spin degeneracies, $k_B$ is the Boltzmann constant, $T$ is the lattice temperature.  The term
\begin{equation}
|C_a(q)|^2=\frac{\hbar}{ 2 \rho v_a} q,
\end{equation}
where the mass density of graphene, $\rho=7.6\times 10^{-8}$ g cm$^{-2}$,  $N_q=\left(\exp(\hbar \omega_q^a/k_B T)-1\right)^{-1}$ is the Bose Einstein distribution function for the phonons and $f(E)=\left(\exp((E-\mu)/k_B T)+1\right)^{-1}$ is the Fermi-Dirac distribution of the electrons.  The scattering matrix element is given by 
\begin{equation}
I^a=\int dS \psi^*_{N',k_y'} V^a_{\bf q} \psi_{N,k_y},
\end{equation}
where $V^a_{\bf q}$ are the electron-phonon coupling matrices for the TA and LA phonons \cite{vonoppen2009,Sohier2014,Mariani2010,Suzuura2002,Manes2007}: 
\begin{equation}
V^{TA}_{\bf q}=e^{i{\bf q}.{\bf r}}\begin{pmatrix}0 & -g_g e^{i2\varphi} \\ g_g e^{-i2\varphi} & 0\end{pmatrix}
\end{equation}
and 
\begin{equation}
V^{LA}_{\bf q}=i e^{i{\bf q}.{\bf r}}\begin{pmatrix}g_d(q) & g_g e^{i2\varphi} \\ g_g e^{-i2\varphi} & g_d(q)\end{pmatrix}.
\label{eq:VLAq}
\end{equation}

The terms $g_g$ and $g_d(q)$ are the electron-phonon coupling matrix elements corresponding to ``gauge''- and ``deformation''-like distortions of the graphene lattice \cite{Sohier2014,Suzuura2002,Manes2007}.   The off-diagonal gauge matrix elements arise from pure shear distortions of the graphene lattice in which the local area of the lattice remains constant and the Fermion couples to the phonon via changes in the local bond lengths.  This type of distortion can be described by a ``synthetic'' gauge field in the Dirac equation \cite{Katsnelsonbook}. It has the effect of changing the position of the Dirac point in the Brillouin zone and is unaffected by screening.  The matrix elements $g_g$ have been estimated using density functional theory (DFT) to have a value in the range $1.5-4.5$ eV \cite{Katsnelsonbook}. In our model, we obtain a good fit to the data with $g_g=4$ eV.   The diagonal matrix elements, $g_d(q)$ arise from deformations of the graphene lattice whereby local areas of the lattice change in size.  These terms shift the energy of the Dirac point. They result in local redistributions in the charge density and are consequently affected by electron screening in the layer and also by the dielectric environment of the graphene layer.   The Thomas-Fermi screening of the deformation electron-phonon coupling matrix element for a phonon with wavevector $q$ is given by
\begin{equation}
g_d(q)={\tilde g_d}/\varepsilon(q).
\end{equation}
Here  ${\tilde g_d}=25$ eV \cite{Efetov2010} is the ``bare'' unscreened electron-phonon coupling constant,
\begin{equation}
\varepsilon(q)=\varepsilon_r\left(1+\frac{q_{tf}}{q}\right)
\end{equation}
and $q_{tf}=4e^2\sqrt{n_s\pi}/(4\pi \hbar \varepsilon_0 \varepsilon_r  v_F)$ is the inverse Thomas-Fermi screening radius. This takes into account screening by the dielectric environment of the graphene layer with dielectric constant, $\varepsilon_r$, and by the electronic charge in the graphene layer  \cite{Hwang2007,Katsnelsonbook}.  When $\varepsilon_r=1$, i.e. for free-standing graphene, $q_{t}\sim8k_F$. Therefore, assuming that on resonance, $q\sim 2k_F$, $\varepsilon(q)\sim5$ and the deformation potential is strongly suppressed.  The $g_d(q)$ term is further screened for graphene on a substrate or when encapsulated by hBN.   Therefore, the TA phonons are unaffected by screening but, in contrast, the on-diagonal parts of the coupling matrix for the LA phonon can be strongly suppressed by screening.  

Evaluating the summations over $k_y$, $k_y'$ and converting the sum over ${\bf q}$ to an integral in polar coordinates,  we obtain the relation for the magnetoconductivity
 \begin{widetext}
 \begin{equation}
\sigma^a_{xx}=\frac{e}{4\pi^2 B k_B T \rho v_a^2} \sum_{N,N'}  \int\int d\varphi dq \ q^4 \sin^2(\varphi) N_q(N_q+1) 
\left[ f(E_{N}-\hbar v_a q) - f (E_{N}) \right] \delta\left(\frac{E_N-E_{N'}}{\hbar v_a}-q \right) |I^{a}_{N,N'}(q,\varphi)|^2.
\label{eq:condint}
\end{equation}
In the high carrier density regime \cite{Kumaravadivel2019} there are no transitions between the conduction and valence band, so that
\begin{equation}
|I^{TA}_{N,N'}(q,\varphi)|^2 = |\frac{i g_g}{2} \left(e^{i2\varphi} \Xi_{N-1,N'} + e^{-i2\varphi} \Xi_{N,N'-1} \right)|^2
\end{equation}
for the TA phonons and 
\begin{equation}
|I^{LA}_{N,N'}(q,\varphi)|^2= |\frac{1}{2} \left[ ig_d (q)\left( \Xi_{N,N'} + \Xi_{N-1,N'-1} \right)  - g_g\left(e^{-i2\varphi} \Xi_{N,N'-1} - e^{i2\varphi} \Xi_{N-1,N'}  \right) \right]|^2
\label{eq:JLA}
\end{equation}
for the LA phonons. Here 
\begin{equation}
\Xi_{N+p,N} =  \left(ie^{-i\varphi}\right)^{p} \sqrt{\frac{N!}{(N+p)!}} \exp\left( -\frac{q^2l^2_B}{4}\right) \left(\frac{ql_B}{\sqrt{2}}\right)^p  L_{N}^{p}\left(\frac{ q^2l_B^2}{2}\right) 
\end{equation}
\end{widetext}
and $L_N^p$ are Laguerre polynomials \cite{Lewis1968}.  The \mbox{magnetoresistivity} components are given by $\rho_{yy}=\sigma_{xx}/(\sigma_{xx}\sigma_{yy}+\sigma_{xy}^2)$ and $\sigma_{xy}=n_se/B$. Under the condition of the experiment \cite{Kumaravadivel2019}, the carrier mobility, $\mu$, is high so that $\mu B\gg 1$ even for fields of a few Tesla.  Hence $\sigma_{xy}\gg\sigma_{xx}\approx\sigma_{yy}$.  By summing the contributions of LA and TA phonon scattering, we then obtain the following relation for the magnetoresistivity:
\begin{equation}
\rho_{yy}=\left(\frac{B}{n_s e}\right)^2 \left(\sigma^{LA}_{xx}+\sigma^{TA}_{xx}\right).
\end{equation}

\section{Discussion}

\begin{figure}[!t]
  \centering
\includegraphics[width=1\linewidth]{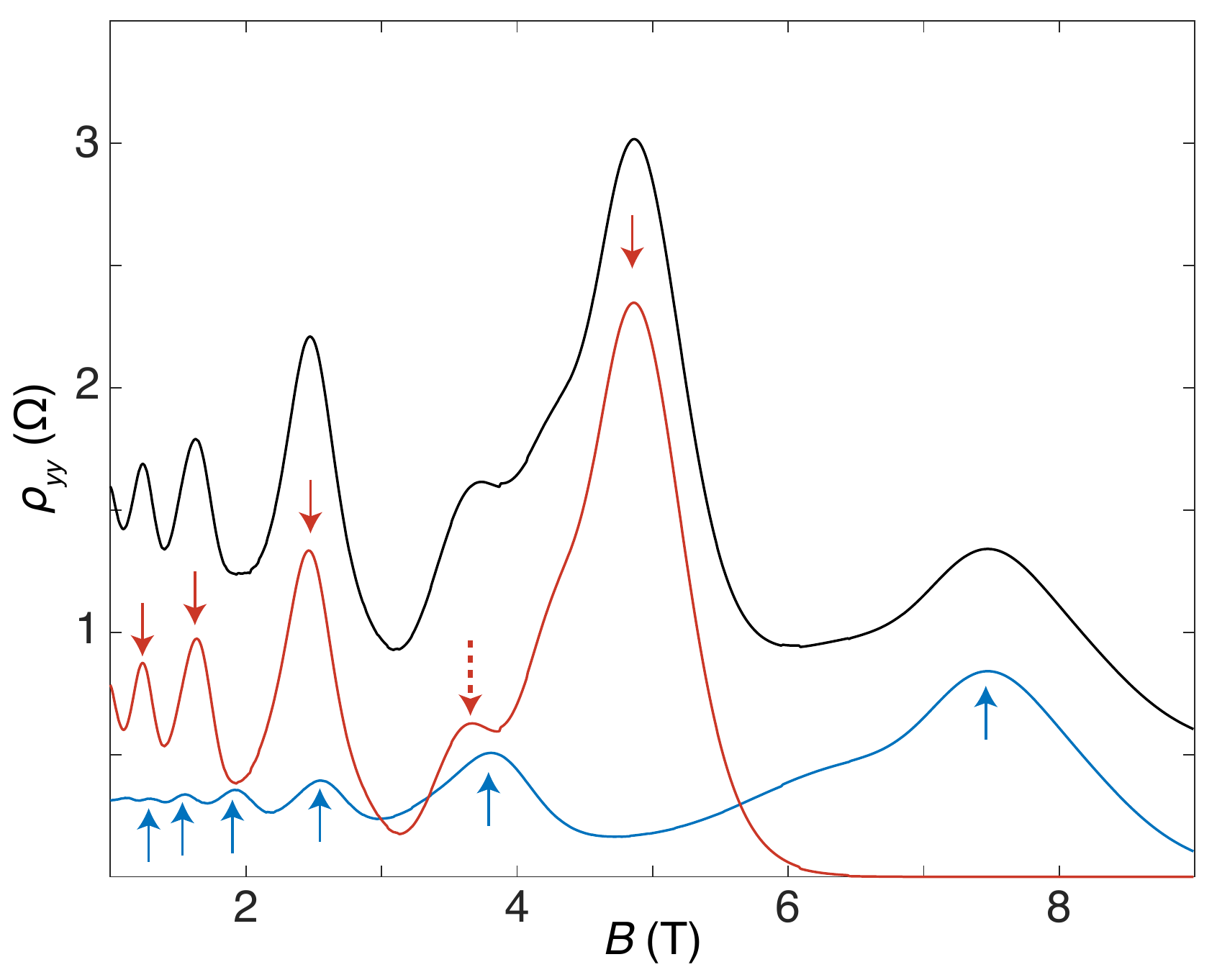}
  \caption{Calculated $\rho_{yy}(B)$ (black) (off-set by 0.5 $\Omega$ for clarity), $\rho^{TA}_{yy}(B)$ (red), and $\rho^{LA}_{yy}(B)$ (blue) with $v_{TA}=13.6$ kms$^{-1}$, $v_{LA}=21.4$ kms$^{-1}$, $n_s=9\times10^{16}$ m$^{-2}$ and $T=70$ K.  Red and blue solid arrows highlight peaks corresponding to the magnetophonon resonance condition in Eq. \ref{eq:Blexact}. Dotted red arrows show additional resonance corresponding to features in $K^{TA}_{n+p,n}(q)$, see Eq. \ref{eq:K}. }
\label{fig:1}
\end{figure}

The black curve in Figure \ref{fig:1}  shows the calculated magnetoresistivity, $\rho_{yy}(B)$. when both TA and LA phonon scattering are included.  The red and blue curves show the separate contributions to $\rho_{yy}(B)$ of the TA and LA phonons respectively, ($\rho^{TA}_{yy}(B)$ and $\rho^{LA}_{yy}(B)$).  First we consider this calculation for free-standing graphene, $\varepsilon_r=1$, at $T=70$ K and $n_s=9\times10^{16}$ m$^{-2}$. Recent DFT calculations \cite{Sohier2014} have estimated the phonon speeds to be $v_{LA}=21.4$ kms$^{-1}$ and $v_{TA}=13.6$ kms$^{-1}$, which we use in our calculation along with\,\cite{CastroNeto2009} $v_F=1.00\times10^6$ ms$^{-1}$.  We find that $\rho_{yy}(B)$ has an oscillatory form and amplitude that corresponds accurately with oscillations observed in recent experiments \cite{Kumaravadivel2019}, see Fig. \ref{fig:ndep}(b).  The maxima in $\rho^{TA}_{yy}(B)$, and $\rho^{LA}_{yy}(B)$, indicated by vertical red and blue arrows, are periodic in $1/B$ and their positions correspond closely to the resonance condition in Eq. \ref{eq:Blexact}.  The plot shows that the contribution of the LA phonons to the total resistivity is relatively weak and appears only as the small peak ($p=1$) in $\rho_{yy}$ at $B\approx7.5$ T.   This is due partly to the suppression of the deformation part of the electron-phonon coupling matrix by electronic screening, see Eq. \ref{eq:VLAq}.  In addition, the energy of the LA phonon is larger than the energy of the TA phonon. Hence there is a lower population of LA phonons than TA phonons at a given temperature.  

Our results support previous theoretical studies of the electron-phonon-induced resistivity in zero magnetic field which show that the contribution to the resistivity by TA phonons is larger than that due to LA phonons \cite{Park2014,Sohier2014}.

We now consider how the overlap integrals of the wavefunctions of the Dirac Fermions lead to small but subtle differences in the magnetophonon resonance condition compared to the semiclassical model based on overlapping cyclotron orbits described in section \ref{sec:2}.

\begin{figure}[!t]
  \centering
\includegraphics[width=1\linewidth]{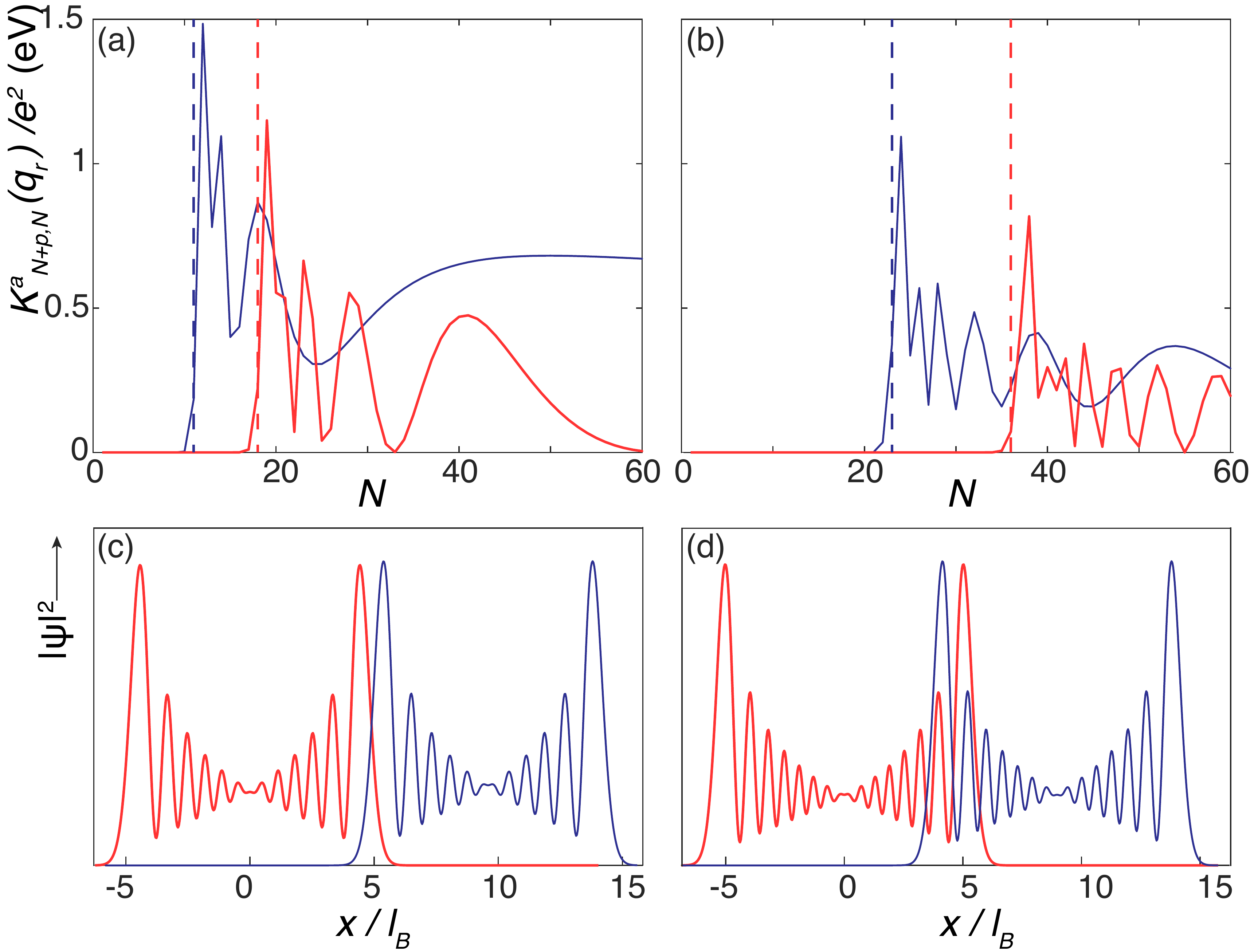}
  \caption{Calculated $K^a_{N+p,N}(q)$ for LA (blue) and TA (red) phonons when $p=1$ (a) and $p=2$ (b).  The vertical dashed lines correspond to the values of $N_p$ determined using the classical cyclotron orbit relation in Eq. \ref{eq:m}.  (c) and (d) show the probability density of the electron wavefunction before (red) and after (blue) it is scattered by a phonon with wavevector $q_y=q_r$, for $N=N_p(a=LA)=11$ (c) and $N=14$ (d).} 
\label{fig:JLAVA}
\end{figure}

These differences illustrate the relaxation of the correspondence principle between classical and quantum mechanics.  Figs. \ref{fig:JLAVA}(a) and (b) show the dependence of the function
\begin{equation}
K^a_{N+p,N}(q_r)=\int d \varphi I^{a}_{N+p,N}(q=q_r,\varphi),
\label{eq:K}
\end{equation}
 on $n$ when $p=1$ and 2 respectively and $q_r=(E_{N+p}-E_N)/\hbar v_a$.  As discussed in our semiclassical analysis (Eq. \ref{eq:m}), the probability of scattering between Landau levels, $K^a_{N+p,N}(q_r)$ is non-zero for $N\gtrsim N_p$ (vertical dashed lines in Fig. \ref{fig:JLAVA}(a) and (b)).   The maximum in $K^a_{N+p,N}(q_r)$ occurs when $N$ is slightly larger than the semiclassical value $N_p$ given by Eq. \ref{eq:m} because the peak in the probability density of the wavefunction does not occur exactly at $r_N$. The maximum overlap between the initial and final states occurs at slightly larger $q$ than the classical estimate in Eq. \ref{eq:kcond},  see plots of $|\Psi_{N+p,X=0}|^2$ (blue) and $|\Psi_{N,X=l_B^2q_r}|^2$ (red) in Fig. \ref{fig:JLAVA}(c).  This leads to a slight systematic deviation between the position of the peaks calculated using Eq. \ref{eq:condint} and that estimated by Eq. \ref{eq:Blexact}.  Therefore the ratios of the phonon speeds to the Fermi velocity deduced from the semiclassical cyclotron orbit relation given by Eq. \ref{eq:Blexact} are slightly lower, by $\sim 5 \%$, than those obtained from the quantum calculation presented in section \ref{sec:kubo}.  
 
For both the LA and TA phonons there are a series of peaks in $K^a_{N+p,N}(q_r)$ for $N>N_p$, see Figs \ref{fig:JLAVA}(a) and (b).  These weaker additional peaks correspond to the overlap of the additional antinodes in the wavefunction for $k<\kappa$, see Fig. \ref{fig:JLAVA}(d).   The red dashed vertical arrows in Fig. \ref{fig:1} highlight additional peaks in $\rho_{yy}^{TA}$ that arise from these extra resonances in $K^{TA}_{N+p,N}(q_r,\varphi)$.  These subtle features do not appear in the experimental data due to LL broadening.

\begin{figure}[!t]
  \centering
\includegraphics[width=1.0\linewidth]{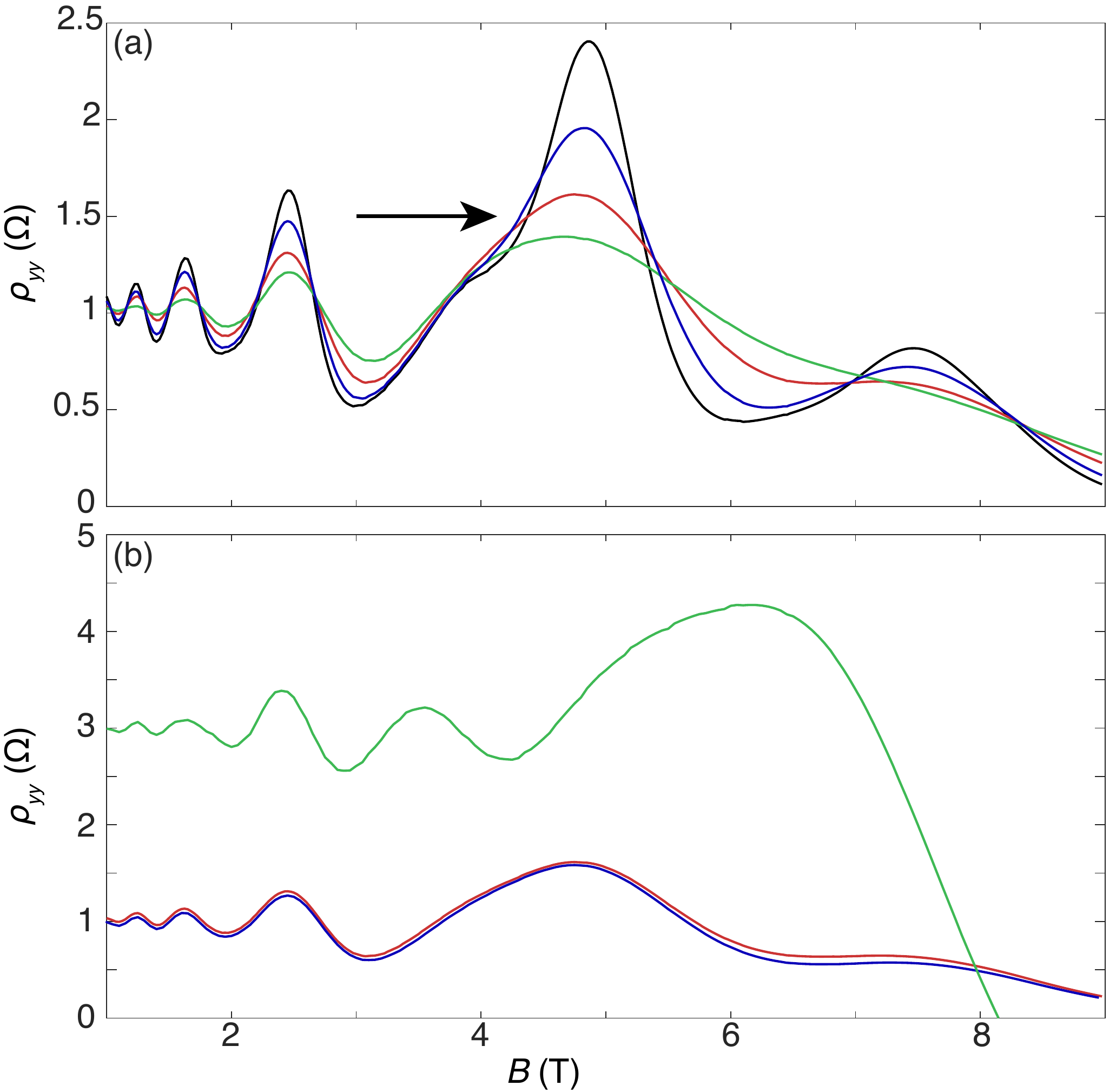}
  \caption{(a) Calculated $\rho_{yy}(B)$ with broadened level carrier distribution when $T=70$ K and $\gamma=0.1$ (black) 0.3 (blue), 0.5 (red) and 0.7 (green) meV.  (b) Calculated $\rho_{yy}(B)$ when $T=70$ K and $\gamma=0.5$ meV with $\varepsilon_r$=1 (blue), $\varepsilon_r=3.5$ (red) and $\varepsilon(q)=1$ (green).  }
\label{fig:broad}
\end{figure}

To model the LL broadening we replace the delta function in Eq. \ref{eq:condint} by 
\begin{equation}
\delta(E) \rightarrow \frac{v_a\hbar}{\Gamma\sqrt{2\pi}}\exp\left(-\frac{E^2}{2\Gamma^2} \right)
\end{equation}
where $E=E_N-E_{N'}-q\hbar v_a$.  We use a Gaussian function to aid convergence of our calculation at high LL indices.   It is known that for the case of elastic short range scattering, for example from charged impurities or defects, the broadening of the LLs is dependent on the square root of magnetic field \cite{Koshino2007,Mori2011,Shon1998,Yang2010}. Therefore we set $\Gamma=\gamma \sqrt{B}$.  

Figure \ref{fig:broad}(a) shows the calculated $\rho_{yy}(B)$ when broadened Landau levels with different values of $\gamma$ are included in our calculation.  For values of $\gamma>0.3$ meV, the secondary resonances in resistivity, which are clearly observed without broadening, see dashed arrows in Fig. \ref{fig:1}, are absent, which is consistent with the measured data.  The secondary resonances, along with the $p=2$ peak for the LA phonon, sum to produce a weak shoulder-like feature of primary peak of $\rho_{yy}(B)$ when $\gamma=0.5$ and 0.7 meV (see horizontal arrow), consistent with the lineshape of the primary peak in the measurements, see horizontal arrow in Fig. \ref{fig:ndep}(b). The best fit to the experimental data is obtained when $\gamma=0.5$ meV, see also Fig. \ref{fig:ndep}(a). Future work could include a more detailed temperature dependent model for LL broadening arising from phonon scattering, see for example Ref. \cite{Funk2015}. This would provide a more accurate fit to the data and would allow a future analysis of the damping out of the oscillations at high temperature.  

Finally, we consider the effect of screening on the magnetoresistance oscillations.  The blue and red curves in Fig. \ref{fig:broad}(b) are plots of $\rho_{yy}(B)$ calculated when $\varepsilon_r=1$ and $\varepsilon_r=3.5$ corresponding to graphene suspended in free-space and graphene encapsulated by boron nitride respectively.  We find that the total resistivity is not strongly dependent on the value of $\varepsilon_r$.  This indicates that screening by carriers in the graphene layer is dominant at these high carrier densities.  We also calculate $\rho_{yy}(B)$ with no screening i.e. $\varepsilon(q)=1$, green curve in Fig. \ref{fig:broad}(b). In this case, the magnetoresistance peak corresponding to LA phonon scattering dominates over that from TA phonon scattering and its position is shifted due to the dominance of the on-diagonal terms in the electron-phonon scattering matrix element.  This is inconsistent with the measurements \cite{Kumaravadivel2019} and highlights the importance of including carrier screening to understand the nature of the measured MPR oscillations.   

\section{Temperature dependence of resistivity in the absence of a magnetic field}

\begin{figure}[t!]
  \centering
\includegraphics[width=1.0\linewidth]{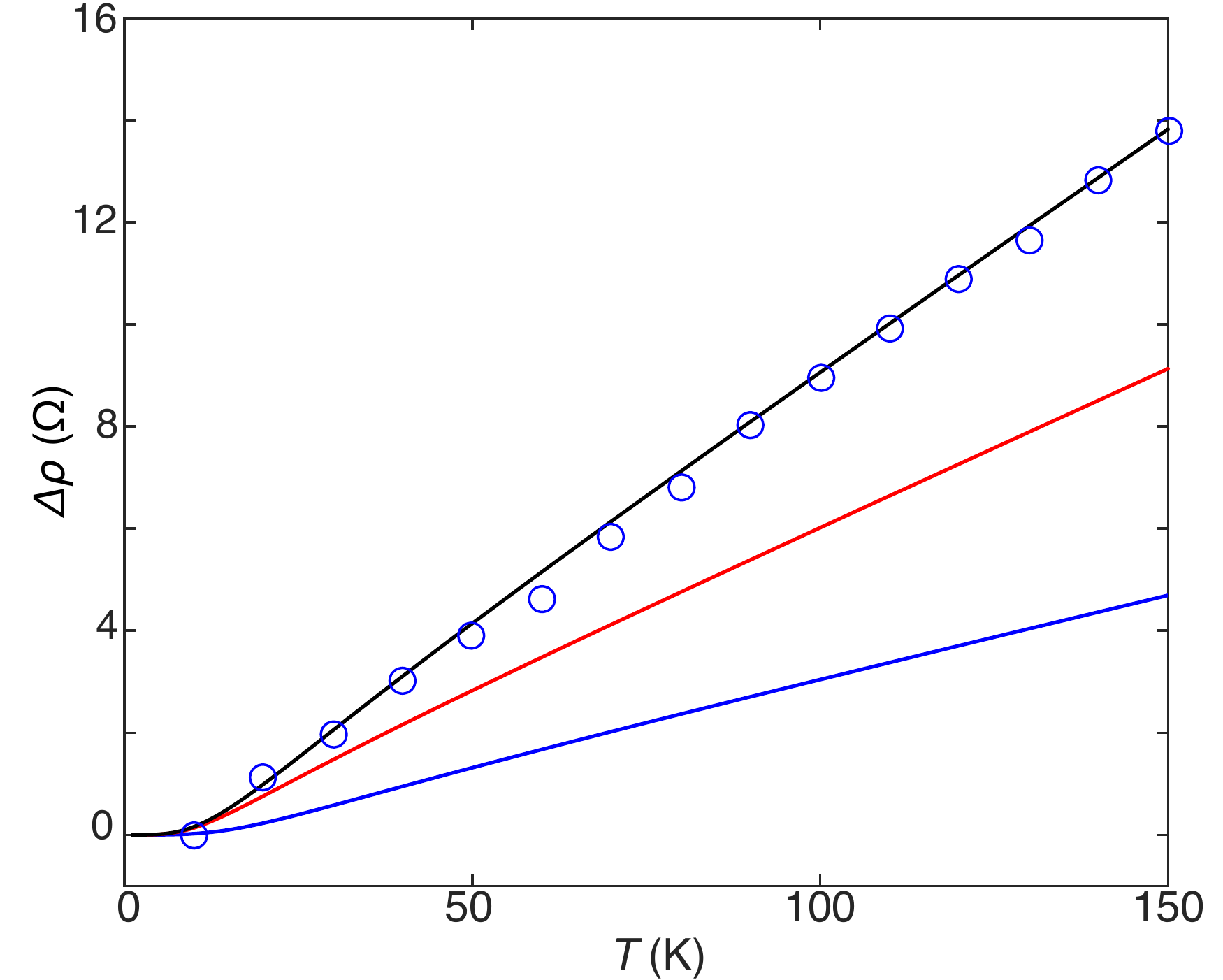}
  \caption{The increase of the resistivity, $\Delta \rho(T)$, with temperature at $B=0$ measured (open blue circles) and calculated (solid black curve) when $n_s=3.2\times10^{12}$ cm$^{-2}$.  The blue and red curves show separately the contribution of the LA and TA phonons respectively to the resistivity, $\rho^{LA}(T)$ and $\rho^{TA}(T)$ }
\label{fig:zeromag}
\end{figure}

This section considers the temperature dependence of the phonon contribution to the resistivity, $\Delta \rho(T)$, of the Hall bar when $B=0$.  To calculate the temperature dependence of the resistivity due to both the TA and LA phonons, we use the linearized Boltzmann equation for temperature-limited resistivity in graphene \cite{Mariani2010}.  The parameters we use in the model are the same as those we use to calculate the form of the MPR oscillations, and we also include the screening of the deformation potential, see section \ref{sec:kubo}.  The black curve in Fig. \ref{fig:zeromag} is the calculated total resistivity when $n_s=3.2\times10^{12}$ cm$^{-2}$ for graphene encapsulated by hBN, and $\epsilon_r=3.5$.  This model agrees quantitatively with the measured dependence of resistivity on temperature, blue open circles, for the graphene sample used in \cite{Kumaravadivel2019}.  The blue and red curves show the contribution to the resistivity of the LA, $\rho^{LA}(T)$ and TA, $\rho^{TA}(T)$, phonons respectively. They reveal that $\rho^{TA}(T)\sim 2 \rho^{LA}(T)$  consistent with the analysis in \cite{Park2014}. This result provides further confirmation of our model parameters and the higher contribution of TA phonon scattering over LA phonon scattering.  It also demonstrates how MPR can be used to elucidate spectroscopically the electron-phonon coupling parameters, which are fundamental to the electronic properties of two-dimensional materials, and which are not accessible by conventional techniques.

\bigskip
\section{Conclusions}

We have presented a model to describe quantitatively the magneto-acoustic phonon resonance recently measured in graphene.  The results of our calculations at low magnetic fields and high carrier densities show a series of oscillations with a form which agree well with the measurements.  This phenomena can be used to determine the speeds of the LA and TA phonons and their relative contributions to the resistivity.   The results thus provide insight into the nature of phonon limited resistivity in graphene.  Previous theoretical work has investigated how high energy optical phonons could give rise to magnetophonon peaks in graphene \cite{Mori2011}.  An experimental realisation would require the generation of hot carriers with energies well in excess of those reported here, either by applying very large electric fields or by optical excitation.


\begin{thebibliography}{9}

\bibitem{Gurevich1961}
Gurevich, V. L. and Firsov, Y. A. On the theory of the electrical conductivity of semiconductors in a magnetic field. {\it J. Exp. Theor. Phys.} {\bf 13}, 137-146 (1961).

\bibitem{Peterson1975}
Peterson, R. L. Chapter 4 The Magnetophonon Effect. in Semiconductors and Semimetals Vol. 10, Pages 221-289 (1975).

\bibitem{Firsov1964}
Firsov, Y. A., Gurevich, V. L., Parfeniev, R. V and Shalyt, S. S. Investigation of a New Type of Oscillations in the Magnetoresistance. {\it Phys. Rev. Lett.} {\bf 12}, 660-662 (1964).

\bibitem{Stradling1968}
Stradling, R. A. and Wood, R. A. The magnetophonon effect in III-V semiconducting compounds. {\it J. Phys. C Solid State Phys.} {\bf 1}, 330 (1968).

\bibitem{Eaves1975}
Eaves, L. et al. Fourier analysis of magnetophonon and two-dimensional Shubnikov-de Haas magnetoresistance structure. {\it J. Phys. C Solid State Phys.} {\bf 8}, 1034-1053 (1975).

\bibitem{Tsui1980}
Tsui, D. C., Englert, T., Cho, A. Y. and Gossard, A. C. Observation of Magnetophonon Resonances in a Two-Dimensional Electronic System. {\it Phys. Rev. Lett.} {\bf 44}, 341-344 (1980).

\bibitem{Nicholas1985}
Nicholas, R. J. The magnetophonon effect. {\it Prog. Quantum Electron.} {\bf 10}, 1-75 (1985).

\bibitem{Dmitriev2012}
Dmitriev, I. A., Mirlin, A. D., Polyakov, D. G. and Zudov, M. A. Nonequilibrium phenomena in high Landau levels. {\it Rev. Mod. Phys.} {\bf 84}, 1709-1763 (2012).

\bibitem{Zudov2001}
Zudov, M. A., Ponomarev, I. V., Efros, A. L., Du, R. R., Simmons, J. A., and Reno, J. L. New class of magnetoresistance oscillations: Interaction of a two-dimensional electron gas with leaky interface phonons. {\it Phys. Rev. Lett.} {\bf 86}, 3614-3617 (2001).

\bibitem{Hatke2009}
Hatke, A. T., Zudov, M. A., Pfeiffer, L. N. and West, K. W. Phonon-induced resistance oscillations in 2D systems with a very high electron mobility. {\it Phys. Rev. Lett.} {\bf 102}, 086808 (2009).

\bibitem{Kumaravadivel2019}
Kumaravadivel, P. et al. Observation of magnetophonon oscillations in extra-large graphene devices. arXiv:1905.00386 (2019).

\bibitem{Hwang2008}
Hwang, E. H. and Das Sarma, S. Acoustic phonon scattering limited carrier mobility in two-dimensional extrinsic graphene. {\it Phys. Rev. B} {\bf 77}, 115449 (2008).

\bibitem{Suzuura2002}
Suzuura, H. and Ando, T. Phonons and electron-phonon scattering in carbon nanotubes. {\it Phys. Rev. B} {\bf 65}, 235412 (2002).

\bibitem{Manes2007}
Ma\~nes, J. L. Symmetry-based approach to electron-phonon interactions in graphene. {\it Phys. Rev. B} {\bf 76}, 045430 (2007).

\bibitem{Mariani2010}
Mariani, E. and von Oppen, F. Temperature-dependent resistivity of suspended graphene. {\it Phys. Rev. B} {\bf 82}, 195403 (2010).

\bibitem{vonoppen2009}
von Oppen, F., Guinea, F. and Mariani, E. Synthetic electric fields and phonon damping in carbon nanotubes and graphene. {\it Phys. Rev. B} {\bf 80}, 075420 (2009).

\bibitem{Sohier2014}
Sohier, T., Calandra, M., Park, C.-H. Bonini, N., Marzari, N., and Mauri, F.  Phonon-limited resistivity of graphene by first-principles calculations: Electron-phonon interactions, strain-induced gauge field, and Boltzmann equation. {\it Phys. Rev. B} {\bf 90}, 125414 (2014).

\bibitem{Park2014}
Park, C. H. et al. Electron-phonon interactions and the intrinsic electrical resistivity of graphene. {\it Nano Lett.} {\bf 14}, 1113-1119 (2014).

\bibitem{Kaasbjerg2012}
Kaasbjerg, K., Thygesen, K. S. and Jacobsen, K. W. Unraveling the acoustic electron-phonon interaction in graphene. {\it Phys. Rev. B} {\bf 85}, 165440 (2012).

\bibitem{Borysenko2010}
Borysenko, K. M., Mullen, J. T., Barry, E. A., Paul, S., Semenov, Y. G., Zavada, J. M., Nardelli Buongiorno, M. and Kim, K. W. First-principles analysis of electron-phonon interactions in graphene. {\it Phys. Rev. B} {\bf 81}, 121412(R) (2010).

\bibitem{Morozov2008}
Morozov, S. V., Novoselov, K. S., Katsnelson, M. I., Schedin, F., Elias, D. C., Jaszczak, J. A. and Geim, A. K. Giant Intrinsic Carrier Mobilities in Graphene and Its Bilayer. {\it Phys. Rev. Lett.} {\bf 100}, 016602 (2008).

\bibitem{Efetov2010}
Efetov, D. K. and Kim, P. Controlling Electron-Phonon Interactions in Graphene at Ultrahigh Carrier Densities. {\it Phys. Rev. Lett.} {\bf 105}, 256805 (2010).

\bibitem{Castro2010}
Castro, Eduardo V., Ochoa, H., Katsnelson, M. I., Gorbachev, R. V., Elias, D. C., Novoselov, K. S., Geim, A. K. and
Guinea, F. Limits on Charge Carrier Mobility in Suspended Graphene due to Flexural Phonons. {\it Phys. Rev. Lett.} {\bf 105}, 266601 (2010).

\bibitem{Alexeev2015}
Alexeev, A. M., Hartmann, R. R. and Portnoi, M. E. Two-phonon scattering in graphene in the quantum Hall regime. {\it Phys. Rev. B} {\bf 92}, 195431 (2015).

\bibitem{Wu2019}
Wu, F., Hwang, E. and Das Sarma, S. Phonon-induced giant linear-in-T resistivity in magic angle twisted bilayer graphene: Ordinary strangeness and exotic superconductivity. {\it Phys. Rev. B} {\bf 99}, 165112 (2019).

\bibitem{Yudhistira2019}
Yudhistira, I. et al. Gauge phonon dominated resistivity in twisted bilayer graphene near magic angle. arXiv:1902.01405 (2019).

\bibitem{Polshyn2019}
Polshyn, H. et al. Phonon scattering dominated electron transport in twisted bilayer graphene. arXiv:1902.00763 (2019).

\bibitem{Jung2015}
Jung, S. et al. Vibrational Properties of h-BN and h-BN-Graphene Heterostructures Probed by Inelastic Electron Tunneling Spectroscopy. {\it Sci. Rep.} {\bf 5}, 16642 (2015).

\bibitem{Vdovin2016}
Vdovin, E. E., Mishchenko, A., Greenaway, M. T., Zhu, M. J., Ghazaryan, D., Misra, A., Cao, Y., Morozov, S. V., Makarovsky, O et al. Phonon-Assisted Resonant Tunneling of Electrons in Graphene-Boron Nitride Transistors. {\it Phys. Rev. Lett.} {\bf 116}, 186603 (2016).

\bibitem{Greenaway2015}
Greenaway, M. T. et al. Resonant tunnelling between the chiral Landau states of twisted graphene lattices. {\it Nat. Phys.} {\bf 11}, 1057-1062 (2015).

\bibitem{Yu2013} 
Yu, G. L. et al. Interaction phenomena in graphene seen through quantum capacitance. {\it Proc. Natl. Acad. Sci.} {\bf 110}, 3282-3286 (2013).

\bibitem{Elias2011}
Elias, D. C. et al. Dirac cones reshaped by interaction effects in suspended graphene. {\it Nat. Phys.} {\bf 7}, 701-704 (2011).

\bibitem{Kubo1965}
R. Kubo, S. J. Miyake, and N. Hashitsume: in Solid State Physics, ed. F. Seitz and D. Turnbull (Academic, New York, 1965) Vol. 17, p. 269

\bibitem{Hamaguchi1990}
Hamaguchi, C. and Mori, N. Magnetophonon resonance in semiconductors. {\it Phys. B Condens. Matter} {\bf 164}, 85-96 (1990).

\bibitem{Mori2011}
Mori, N. and Ando, T. Magnetophonon Resonance in Monolayer Graphene. {\it J. Phys. Soc. Japan} {\bf 80}, 44706 (2011).

\bibitem{Shon1998}
Shon, N. H. and Ando, T. Quantum Transport in Two-Dimensional Graphite System. {\it J. Phys. Soc. Japan} {\bf 67}, 2421-2429 (1998).

\bibitem{McCannbook}
McCann E. (2011) Electronic Properties of Monolayer and Bilayer Graphene. In: Raza H. (eds) Graphene Nanoelectronics. NanoScience and Technology. Springer, Berlin, Heidelberg

\bibitem{CastroNeto2009}
Castro Neto, A. H., Guinea, F., Peres, N. M. R., Novoselov, K. S. and Geim, A. K. The electronic properties of graphene. {\it Rev. Mod. Phys.} {\bf 81}, 109-162 (2009).

\bibitem{Beenakker1991}
Beenakker, C. W. J. and van Houten, H. Quantum Transport in Semiconductor Nanostructures. {\it Thin Solid Films} {\bf 393}, 1-228 (1991).

\bibitem{Katsnelsonbook}
Katsnelson, M.I. {\it Graphene: Carbon in Two Dimensions.}  (Cambridge University Press, 2012).

\bibitem{Hwang2007}
Hwang, E. H. and Das Sarma, S. Dielectric function, screening, and plasmons in two-dimensional graphene. {\it Phys. Rev. B} {\bf 75}, 205418 (2007).

\bibitem{Lewis1968}
Lewis, J. K. and Hougen, J. T. Avoided crossings in bound potential-energy curves of diatomic molecules: Derivation and analysis of the vibrational hamiltonian. {\it J. Chem. Phys.} {\bf 48}, 5329-5336 (1968).

\bibitem{Koshino2007}
Koshino, M. and Ando, T. Diamagnetism in disordered graphene. {\it Phys. Rev. B} {\bf 75}, 235333 (2007).

\bibitem{Yang2010}
Yang, C. H., Peeters, F. M. and Xu, W. Density of states and magneto-optical conductivity of graphene in a perpendicular magnetic field. {\it Phys. Rev. B} {\bf 82}, 205428 (2010).

\bibitem{Funk2015}
Funk, H., Knorr, A., Wendler, F. and Malic, E. Microscopic view on Landau level broadening mechanisms in graphene. {\it Phys. Rev. B} {\bf 92}, 205428 (2015).


\end{thebibliography}
\end{document}